\documentclass[11pt]{article}
\usepackage[square,authoryear]{natbib}
\usepackage{marsden_article}

\renewcommand{\todo}[1]{}

\begin{document}

\title{\LARGE \bf A Class of Integrable Geodesic Flows on the Symplectic Group
and the Symmetric Matrices}

\author{Anthony M. Bloch\thanks{
            Research partially supported by the NSF.}
            \\Department of Mathematics
            \\ University of Michigan \\ Ann Arbor MI 48109
            \\{\small abloch@math.lsa.umich.edu}
            \and
Arieh Iserles
            \\Department of Applied Mathematics and Theoretical Physics
            \\University of Cambridge
            \\ Wilberforce Road
            \\Cambridge CB3 0WA
            \\{\small A.Iserles@damtp.cam.ac.uk}
\and
Jerrold E. Marsden\thanks{
Research partially supported by the
California Institute of Technology
and NSF.}
\\Control and Dynamical Systems 107-81
\\ California Institute of Technology
\\ Pasadena, CA 91125
\\ {\small marsden@cds.caltech.edu}
             \and
Tudor S. Ratiu\thanks{
Research partially supported by
the Swiss NSF.}\\
Section de Math\'ematiques\\
            Ecole Polytechnique F\'ed\'erale de
Lausanne \\
CH-1015 Lausanne, Switzerland
\\{\small Tudor.Ratiu@epfl.ch}
}

\date{\small This version: April 19, 2006}
\maketitle

\begin{abstract} 
This paper shows that the left-invariant geodesic flow  on the symplectic group relative to the Frobenius metric is an integrable system that is not contained in the Mishchenko-Fomenko class of rigid body metrics. This system may be expressed as a flow on symmetric matrices and is bi-Hamiltonian. This analysis is extended to cover flows on symmetric matrices when 
an isomorphism with the symplectic Lie algebra does not 
hold. The two Poisson structures associated with this system, including an analysis of its Casimirs, are completely analyzed.
Since the system integrals are not generated by its Casimirs it is 
shown that the nature of integrability is fundamentally different 
from that exhibited in the Mischenko-Fomenko setting.
\end{abstract}

\section{Introduction}
This paper continues the analysis, begun in \cite{BlIs2005},  of the set of ordinary
differential equations 
\begin{equation}
  \label{eq:1.1}
   \dot{X } = [X^2, N],
\end{equation}
where $X \in  \operatorname{Sym} (n)$, the linear space of $n \times n$ symmetric
matrices,  $\dot{X}$ denotes the time derivative, $N \in \mathfrak{so}(n)$, the
space
of skew symmetric $n \times n $ matrices, is given, and where initial conditions
$X(0)=X_0 \in \operatorname{Sym}(n)$ are also given. 

It is easy to check that $[X^2, N] \in \operatorname{Sym} (n)$, so that if the
initial condition is in $\operatorname{Sym} (n)$, then 
$X(t)\in\operatorname{Sym}(n)$ for all $t $. Also, because of the straightforward
identity $ \left[ X ^2, N \right] = \left[X, XN + NX \right]$, this equation may be
rewritten in the Lax form
\begin{equation}
  \label{eq:1.2}
 \dot{X} = [X,XN+NX],
\end{equation}
again with initial conditions $X(0)=X_0\in\operatorname{Sym}(n)$.

We show below that this system may be viewed as a Lie-Poisson system on the dual of the symplectic Lie algebra if $N$ invertible, in which case it is geodesic, and on the dual of a more 
general Lie algebra on symmetric matrices for arbitrary $N$.  The system is bi-Hamiltonian and is not in the Mischenko-Fomenko
class of integrable (geodesic) rigid body systems (\cite{MiFo1976}).
Despite this, we prove that it is integrable on the generic symplectic leaf of the corresponding phase space if $N $ is invertible or of nullity one. We use the Lax pair with parameter found
in \cite{BlIs2005} to find a class of integrals
that we show are in involution using the bi-Hamiltonian structure
and the technique of \cite{MoPi1996}. Independence is  proved directly since the method in \cite{MiFo1976} does not apply to this system, even though the  integrals are obtained from the Casimirs with a shifted argument.  Indeed, this system appears to fundamentally different from completely integrable systems either of rigid body or Toda type (on symmetric matrices).

If $N $ is not invertible, there is no isomorphism of the Lie algebra induced by $N $ with the symplectic Lie algebra. We extend our analysis  of the system to this case and study the Poisson geometry of the dual of this Lie algebra determining the generic leaves and the Casimir functions of both Poisson structures relative to which the system \eqref{eq:1.1} is bi-Hamiltonian.

We want to emphasize that the system \eqref{eq:1.1} (or \eqref{eq:1.2}) for $N $ invertible is thus \textit{a new integrable geodesic flow} of a left invariant metric on the Lie group $\operatorname{Sp}(n, \mathbb{R}) $. So far the only known left invariant metrics whose geodesic flows are integrable on the Lie group $\operatorname{Sp}(n, \mathbb{R})$ are the rigid body metrics of  \cite{MiFo1976}. Finding integrable geodesic flows on Lie groups for left invariant metrics that are not of rigid body type is a daunting task. System \eqref{eq:1.1} is the only one known to us on any semisimple Lie algebra with the exception of $\mathfrak{so}(4)$, where we review the stituation below.

Even for the case of $\operatorname{SO}(4)$ there is only one  known geodesic flow that is not of rigid body type (see \cite{Mishchenko1970, Manakov1976, MiFo1976, Ratiu1980} for the definition of such metrics). There are 
three integrable cases of left invariant metrics for geodesic flow
on SO(4): the metric used in \cite{Manakov1976} (which goes to the Clebsch case by contraction to the Euclidean group) and two other cases that correspond to left invariant metrics that are not diagonal in the standard basis of $\mathfrak{so}(4)$. The first one is obtained from deformation of the classical Lyapunov-Steklov integrable case on $\operatorname{SE}(3)$ by deforming the Lie algebra $\mathfrak{se}(3)$ to $\mathfrak{so}(4)$; the integrability of the corresponding system is due to \cite{BoMaSo2001}. The last case has a fourth
quartic constant of the motion and is the genuinely new integrable geodesic case found by \cite{AdvM1986}; a $\mathfrak{g}_2$ Lax pair for this system was given in \cite{ReSe1986}. \cite{Sokolov2001} showed that these two cases are
not linearly equivalent. The rigid body metric used by \cite{Manakov1976} is the only algebraic completely integrable case for a left invariant metric that is diagonal in the standard basis of $\mathfrak{so}(4)$ (\cite{AdvM1982, Haine1984}). The state of the art regarding these systems is contained in Theorem 8.3, page 270, of \cite{AdvMVa2004}: in a certain large class of metrics (non-degenerate half-diagonal metrics with some weight homogeneity
conditions) these three cases are the only algebraically completely
integrable geodesic flows.  Whether these three cases are the only
algebraically completely integrable geodesic flows in the class of
\textit{all} left invariant metrics is still an open  question. See
\cite{Sokolov2002} for a review and references of what is known about a related system, the Kirchhoff case of the motion of a rigid body in an ideal fluid.

The structure of the paper is as follows. In Section
\ref{LieAlgebra} we consider the Lie algebra structure on symmetric
matrices induced by $N$ and the special case of the isomorphism to 
$\mathfrak{sp}( \mathbb{R}^n)$. In Section \ref{EP} we 
analyze the bi-Hamiltonian structure as well as the symplectic leaves and 
Casmirs of both structures. In Section \ref{Sec} we compare our 
system with the sectional operator systems of Mischenko and Fomenko and conclude that \eqref{eq:1.1} is not in this family, thereby showing that it is a new geodesic flow that is not of rigid body type on the Lie group $\operatorname{Sp}(n \mathbb{R})$.
In Section \ref{Lax} we analyze the Lax pair with parameter and find a family with the right number of integrals of motion that is a candidate for Liouville integrability. In Section \ref{Involution}
we prove involution of the integrals using the bi-Hamiltonian 
structure. In Section \ref{sec:independence} we analyze indepedence 
and finally we discuss some future work in Section \ref{Conclusion}.

\section{The Lie Algebra and the Euler-Poincar\'e Form}
\label{LieAlgebra}
We can regard $N$ as a Poisson tensor on $\mathbb{R}^n$ by defining the bracket of
two functions $f , g$ as
\begin{equation} \label{NPB}
\left\{ f, g \right\} _N =  (\nabla f ) ^T N \nabla g. 
\end{equation}

The Hamiltonian vector field associated with a function $h$ (with the 
convention that $\dot{f} (z) =  X _h (z) \cdot \nabla f(z) = \left\{f, h
\right\} (z) $) is given by
\begin{equation} \label{NHamvf}
X _h (z) = N \nabla h (z),
\end{equation}
as is easily checked.

For each $X \in \operatorname{Sym}(n)$ define the quadratic Hamiltonian $Q_X$
by
\[
Q _X (z) := \frac{1}{2} z ^T X z, \quad z \in \mathbb{R}^n.
\]
Let $\mathcal{Q} : = \{Q_X \mid X \in \operatorname{Sym}(n) \}$ be the vector
space of all such functions. Note that  the map $Q : X \in
\operatorname{Sym}(n)
\mapsto Q_X \in \mathcal{Q}$ is an isomorphism. 

Using \eqref{NHamvf} it follows that the Hamiltonian vector field of
$Q _X $ has the form
\begin{equation} \label{NQHamvf}
X _{Q _X} (z) = N X z.
\end{equation}

Next, we compute the Poisson bracket of two such quadratic functions.

\begin{lemma} For $X, Y \in \operatorname{Sym}(n)$, we have
\begin{equation} \label{PBQHam}
\left\{ Q _X, Q _Y \right\} _N =  Q _{ [X, Y] _N },
\end{equation}
where $[X, Y] _N = XNY - Y N X \in \operatorname{Sym} (n)$. In addition,
$\operatorname{Sym}(n)$ is a Lie algebra relative to the Lie bracket $[\cdot,
\cdot ]_N $. Therefore, $Q : X \in (\operatorname{Sym}(n), [ \cdot , \cdot
]_N) \mapsto Q_X \in ( \mathcal{Q}, \{ \cdot , \cdot \}_N) $ is a Lie algebra
isomorphism.
\end{lemma}

\begin{proof}
Using \eqref{NPB} we have
\begin{align*}
\left\{ Q _X, Q _Y \right\} _N (z) 
& =  \left( \nabla Q _X \right) (z) ^T N \left( \nabla Q _Y \right)(z)  
=  \left( X z \right) ^T N Y z \
=  z^TXNY z \\
& =  \frac{1}{2} z ^T \left( X N Y - Y N X \right) z 
=  Q _{[X,Y ] _N } (z).
\end{align*}
Recall that the notation $Q _V$ is reserved only for symmetric matrices $V$.
Since $X, Y \in \operatorname{Sym}(n) $ implies that $[X, Y] _N = XNY - Y N
X \in \operatorname{Sym}(n)$ we can write $Q _{[X,Y ] _N } $ in the
preceding equation.

The bracket $[ \cdot , \cdot ]_N $ on $\operatorname{Sym}(n) $ is clearly
bilinear and  antisymmetric. The Jacobi identity is a straightforward direct
verification.
\end{proof}

It is a general fact that Hamiltonian vector fields and  Poisson brackets are 
related by
\begin{equation} \label{HVFPB}
\left[ X _f, X _g \right] = - X _{ \left\{ f, g \right\} },
\end{equation} 
where the bracket on the left hand side is the Jacobi-Lie bracket. Thus, it 
is natural to look at the corresponding algebra of Hamiltonian vector fields
on the Poisson manifold $(\mathbb{R}^n, \{ \cdot , \cdot \}_N)$ associated to
quadratic Hamiltonians. If we take
$f = Q _X$ and $g = Q _Y$, with $X _f = NX $ and $X _g =NY $, and recall that
the Jacobi-Lie bracket of {\it linear} vector fields is the negative of the
commutator of the associated matrices, then we have the following result.
\begin{proposition}
Equations \eqref{PBQHam} and \eqref{HVFPB}
imply
\begin{equation}
\label{Niso}
N [X, Y ] _N =[NX, NY ]\,.
\end{equation}
\end{proposition}
This can, of course, be easily verified by hand.

Letting $\mathcal{LH}$ denote
the Lie algebra of linear Hamiltonian vector fields
on $\mathbb{R}^n $ relative to the commutator bracket of matrices,
\eqref{Niso} states that the map 
$$X \in (\operatorname{Sym} (n), [ \cdot ,
\cdot ]_N) \mapsto NX \in ( \mathcal{LH}, [ \cdot , \cdot ])$$
is a
homomorphism of Lie algebras\footnote{We thank Gopal Prasad for suggesting
isomorphisms of this type; they are closely related to well-known properties
of linear Hamiltonian vector fields, as in \cite{MaRa1994}, Proposition
2.7.8.}.  

\paragraph{Invertible Case.} If $N$ is invertible, then  this homomorphism is an
isomorphism.
In addition, the non-degeneracy of $N$ implies that $n$ is 
even and that $\mathbb{R}^n$ is a symplectic vector space relative to the
symplectic form defined by $N ^{-1}$. Therefore, the Lie algebra
$(\mathcal{LH}, [ \cdot ,
\cdot ])$ is isomorphic to the  Lie  algebra $\mathfrak{sp}( \mathbb{R}^n,
N ^{-1})$ of linear symplectic maps of $\mathbb{R} ^n$ relative to the
symplectic form $N ^{-1}$, that is, to the classical Lie algebra
$\mathfrak{sp}(n, \mathbb{R})$. Note that this means that $(NX)^TN ^{-1} + N
^{-1}(NX) = 0$.

We summarize these considerations  in the following  statement.

\begin{proposition} 
\label{sym_isomorphism}
Let $N \in \mathfrak{so}(n)$. The map $Q : X \in (\operatorname{Sym}(n), [
\cdot , \cdot ]_N) \mapsto Q_X \in ( \mathcal{Q}, \{ \cdot , \cdot \}_N) $ is
a Lie algebra isomorphism. The map $X \in (\operatorname{Sym} (n), [
\cdot ,
\cdot ]_N) \mapsto NX \in ( \mathcal{LH}, [ \cdot , \cdot ])$ is a Lie algebra
homomorphism and if $N $ is invertible it induces an isomorphism of
$(\operatorname{Sym} (n), [ \cdot ,
\cdot ]_N)$ with $\mathfrak{sp}(n, \mathbb{R})$.
\end{proposition}

\paragraph{The Euler-Poincar\'e Form}
The Euler-Poincar\'e form for the equations can be derived as follows.  
Identify $\operatorname{Sym} (n)$ with its dual using the the positive
definite inner  product 
\begin{equation}
\label{sym_inner_product}
\left\langle \! \left\langle X, Y \right\rangle \!
\right\rangle := \operatorname{trace}\left(XY\right), \quad \text{for} \quad
X, Y
\in
\operatorname{Sym}(n).
\end{equation}
\paragraph{Remark.} The inner product $\left\langle\!\left\langle X, Y
\right\rangle\!\right\rangle$ is not $\operatorname{ad}$-invariant relative to the
$N$-bracket,
but another one, namely $\kappa _N (X,Y) := \operatorname{trace} (NX NY) $ is
invariant, as is easy to check. \medskip

Define the Lagrangian  $l: \operatorname{Sym} (n) \rightarrow \mathbb{R}$ on
the Lie algebra $(\operatorname{Sym} (n), [ \cdot , \cdot ]_N)$ by
\begin{equation}
l(X)=
\frac{1}{2}\operatorname{trace}  \left(X^2\right)=
\frac{1}{2}\operatorname{trace}  \left(  XX^T\right)=: 
\frac{1}{2} \left\langle \! \left\langle X,X \right\rangle \! \right\rangle.
\label{keylagrangian}
\end{equation}

\begin{proposition}
The equations 
\begin{equation}
\dot{X}=[X^2,N]
\end{equation}
are the Euler-Poincar\'e equations\footnote{For a general discussion of the 
Euler-Poincar\'e equations, see, for instance,
\cite{MaRa1994}.} corresponding to the Lagrangian
{\rm \eqref{keylagrangian}} on the Lie algebra $(\operatorname{Sym} (n),
[\cdot ,\cdot ]_N)$.
\end{proposition}

\begin{proof}
Recall that the general  (left) Euler-Poincar\'e equations on a Lie algebra 
$\mathfrak{g}$ associated with a Lagrangian $l: \mathfrak{g} \rightarrow
\mathbb{R}$ are given by
\begin{equation*}
\frac{d}{dt}Dl( \xi) = \operatorname{ad}_\xi ^{\ast} Dl( \xi), 
\end{equation*}
where $Dl( \xi) \in \mathfrak{g}^\ast$ is the Fr\'echet derivative of $l $ at
$\xi$. Equivalently, for each fixed $\eta \in \mathfrak{g}$, we
have
\begin{equation} 
\label{epalt}
\frac{d}{dt} D l (\xi) \cdot \eta = D l (\xi) \cdot [ \xi, \eta ].
\end{equation}
In our case, letting $\xi = X $ and $\eta = Y $ arbitrary, time-independent,
equations \eqref{epalt} become
\begin{align*}
\frac{d}{dt} \left\langle \! \left\langle X,Y \right\rangle \! \right\rangle 
&=\left\langle \! \left\langle  X,[X,Y]_N \right\rangle \! \right\rangle \\
&=\left\langle \! \left\langle  X,XNY-YNX \right\rangle \! \right\rangle ;
\end{align*}
that is,
\begin{align*}
\operatorname{trace} \left( \dot{X}  Y \right)  & =
\operatorname{trace} \left( X (XNY - Y NX ) \right) \\
& = \operatorname{trace} \left( (X ^2 N - NX ^2) Y \right),
\end{align*}
which gives the result.
\end{proof}

\paragraph{General Case--Noninvertible $N$.} We next determine the structure of the
Lie algebra $(\operatorname{Sym}(n), [\cdot, \cdot]_N)$ for a 
general skew-symmetric matrix $N$. The point of departure is the fact that if $N$ is
nondegenerate, 
then
$X \in (\operatorname{Sym} (n), [\cdot , \cdot ]_N) \mapsto NX \in ( \mathcal{LH}, [
\cdot , \cdot ]) = 
(\mathfrak{sp}( \mathbb{R}^n, N^{-1}), [ \cdot , \cdot ])$ is a Lie algebra
isomorphism. Recall that if $
\mathbb{R}^n$ has an inner product, which we shall take in what follows to be the
usual dot product 
associated to the basis in which the skew-symmetrix matrix $N$ is given, and $L:
\mathbb{R}^n 
\rightarrow \mathbb{R}^n$ is a linear map, then $\mathbb{R}^n$ decomposes
orthogonally as $
\mathbb{R}^n = \operatorname{im} L^T   \oplus   \ker L $. Taking $L=N$ in this
statement and recalling 
that $N^T = - N$, we get the orthogonal decomposition  $\mathbb{R}^n = 
\operatorname{im} N \oplus \ker N
$. Let $2p = \operatorname{rank}N$ and $d: = n-2p$. Then $\bar{N}: = N|_
{\operatorname{im} N}: \operatorname{im} N \rightarrow \operatorname{im} N$
defines a
nondegenerate skew symmetric bilinear form and, by the previous proposition,
$(\operatorname{Sym} 
(2p), [\cdot , \cdot ]_{\bar{N}})$ is isomorphic as a Lie algebra to $(\mathfrak{sp}(
\mathbb{R}^{2p}, 
\bar{N}^{-1}), [ \cdot , \cdot ])$. In this direct sum decomposition of
$\mathbb{R}^n$, the skew-
symmetric matrix $N$ takes the form
\[
N = \begin{bmatrix}
\bar{N} & 0  \\
0 & 0  
 \end{bmatrix},
\]
where $\bar{N}$ is a $(2p) \times (2p)$ skew-symmetric nondegenerate matrix.

The Lie algebra $(\operatorname{Sym} (2p), [\cdot , \cdot ]_{\bar{N}})$ acts on the
vector space $
\mathcal{M}_{(2p) \times d}$ of $(2p) \times d$ matrices (which we can think of as
linear maps of $\ker N $ to $\operatorname{im} N $) by $S \cdot A : = S \bar{N} A$,
where $S \in (\operatorname{Sym}(2p), [\cdot \, \cdot ]_{\bar{N}})$ and $A \in
\mathcal{M}_{(2p) \times d}$. Indeed, if $S, 
S' \in \operatorname{Sym} (2p)$ and $A \in \mathcal{M}_{(2p) \times d}$, then 
\begin{align} \label{SAction}
[S,S']_{\bar{N}} \cdot A &= (S\bar{N}S' - S' \bar{N} S)\bar{N} A
= S\bar{N}S'\bar{N} A - S' \bar{N} S \bar{N} A \nonumber \\
&= S \cdot (S' \cdot A) - S' \cdot (S \cdot A).
\end{align}
Now form the semidirect product $\operatorname{Sym} (2p)
\,\circledS\,\mathcal{M}_{(2p)\times d}$. 
Its bracket is defined by
\begin{align} \label{SDBracket}
[(S, A), (S', A')] &= ([S,S']_{\bar{N}}, S\cdot A' - S' \cdot A)
\nonumber \\
& = (S\bar{N} S' - S' \bar{N} S, S \bar{N} A' - S' \bar{N} A)
\end{align}
for any $S, S' \in \operatorname{Sym} (2p)$ and $A, A' \in \mathcal{M}_{(2p)\times
d}$. 

Next, define the $\operatorname{Sym}(d)$-valued Lie algebra two cocycle 
$$C: \operatorname{Sym} (2p) 
\,\circledS\,\mathcal{M}_{(2p)\times d} \times \operatorname{Sym} (2p)
\,\circledS\,\mathcal{M}_{(2p)
\times d} \rightarrow \operatorname{Sym}(d)$$ 
by
\begin{equation} \label{SDcocycle}
C((S, A), (S', A')) : = A^T \bar{N} A' - (A')^T \bar{N} A
\end{equation}
for any $S, S' \in \operatorname{Sym} (2p)$ and $A, A' \in \mathcal{M}_{(2p)\times
d}$. The cocycle 
identity 
\begin{align*}
&C([(S, A), (S', A')], (S'', A'')) + C([(S', A'), (S'', A'')], (S, A))
\nonumber \\
& \qquad  + C([(S'', A''), (S, A)], (S', A')) = 0
\end{align*}
for any $S, S', S'' \in \operatorname{Sym} (2p)$ and $A, A', A'' \in
\mathcal{M}_{(2p)\times d}$ is a 
straightforward verification.
Now extend $\operatorname{Sym} (2p) \,\circledS\,\mathcal{M}_{(2p)\times d}$ by this
cocycle. That is, form the vector space $(\operatorname{Sym} (2p)
\,\circledS\,\mathcal{M}_{(2p)\times d} )\oplus 
\operatorname{Sym}(d)$ and endow it with the bracket
\begin{equation} \label{ExtendedSDBracket}
[(S, A, B), (S', A', B')]^C : = (S\bar{N} S' - S' \bar{N} S, S \bar{N} A' - S'
\bar{N} A, A^T \bar{N} A' - (A')^T 
\bar{N} A)
\end{equation}
for any $S, S' \in \operatorname{Sym} (2p)$, $A, A' \in \mathcal{M}_{(2p)\times d}$,
and $B, B' \in 
\operatorname{Sym}(d)$.

\begin{proposition} \label{SymStructure}
The map 
$$
\Psi: ((\operatorname{Sym} (2p) \,\circledS\,\mathcal{M}_{(2p)\times d} )\oplus
\operatorname{Sym} (d), [\cdot, \cdot]^C) \rightarrow (\operatorname{Sym} (n),
[\cdot, \cdot]_N)
$$ 
given by
\begin{equation}
\Psi(S, A, B) := 
\begin{bmatrix}
S & A  \\
A^T & B  
 \end{bmatrix}
\end{equation}\
is a Lie algebra isomorphism.
\end{proposition}

\begin{proof} 
It is obvious that $\Psi$ is a vector space isomorphism so only the Lie algebra
homomorphism 
condition needs to be verified. So, let $(S, A, B), (S', A', B') \in 
(\operatorname{Sym} (2p) \,\circledS\,
\mathcal{M}_{(2p)\times d} )\oplus \operatorname{Sym}(d)$ and compute
\begin{align*}
&\Psi([(S, A, B), (S', A', B')]) = 
\Psi(S\bar{N} S' - S' \bar{N} S, S \bar{N} A' - S' \bar{N} A, A^T \bar{N} A' - (A')^T
\bar{N} A) \\
& \qquad =
\begin{bmatrix}
S\bar{N} S' - S' \bar{N} S & S \bar{N} A' - S' \bar{N} A  \\
(S \bar{N} A' - S' \bar{N} A)^T & A^T \bar{N} A' - (A')^T \bar{N} A  
 \end{bmatrix} \\
& \qquad = 
\begin{bmatrix}
S & A  \\
A^T & B  
 \end{bmatrix}
\begin{bmatrix}
\bar{N} & 0  \\
0 & 0  
 \end{bmatrix}
 \begin{bmatrix}
S' & A'  \\
(A')^T & B'  
 \end{bmatrix}
 -  \begin{bmatrix}
S' & A'  \\
(A')^T & B'  
 \end{bmatrix}
 \begin{bmatrix}
\bar{N} & 0  \\
0 & 0  
 \end{bmatrix}
 \begin{bmatrix}
S & A  \\
A^T & B  
 \end{bmatrix} \\
 & \qquad = [\Psi(S,A,B), \Psi(S',A',B')]_N
\end{align*}
as required.
\end{proof}

\section{Poisson Structures}
\label{EP}

Identifying $\operatorname{Sym}(n)$ with its dual using  the  inner product
\eqref{sym_inner_product} endows $\operatorname{Sym}(n)$ with the the (left,
or minus) Lie-Poisson bracket
\begin{equation} \label{LiePB}
\left\{ f, g \right\} _N (X) = 
- \operatorname{trace}  \Big[ X \Big( \nabla f (X) N \nabla g (X) -
\nabla g (X) N \nabla f (X) \Big) \Big],
\end{equation}
where $\nabla f $ is the gradient of $f$ relative to the inner product
$\left\langle \! \left\langle \cdot  , \cdot  \right\rangle \!\right\rangle $
on $\operatorname{Sym} (n)$. It is easy to check that the  equations
$\dot{X} = \left[X^2, N \right]$ are Hamiltonian relative to the function $l
$ defined  in \eqref{keylagrangian} and the Lie-Poisson bracket
\eqref{LiePB}. 

\bigskip

Later on we shall also need the {\it frozen} Poisson bracket 
\begin{equation} \label{FLiePB}
\left\{ f, g \right\} _{FN} (X) = - \operatorname{trace}\Big(\nabla
f(X)N\nabla g (X) - \nabla g (X) N \nabla f (X) \Big).
\end{equation}
It is a general fact that the Poisson structures \eqref{LiePB} and 
\eqref{FLiePB} are \emph{compatible} in the sense that their sum is a Poisson
structure (see e.g. Exercise 10.1-5 in \cite{MaRa1994}).

For what follows it is important to compute the Poisson tensors corresponding
to the above Poisson brackets. Recall that the Poisson tensor can be viewed
as a vector bundle morphism $B : T^*(\operatorname{Sym}(n)) \rightarrow
T(\operatorname{Sym}(n))$ covering the identity. It is defined by $B(
\mathbf{d} h) = \{\cdot , h\}_N$ for any locally defined smooth function
$h$ on $\operatorname{Sym}(n)$. Since $\operatorname{Sym}(n)$ is a vector
space, these bundles are trivial and hence the value $B_X $ at $X \in
\operatorname{Sym}(n)$ of the Poisson tensor $B$ is a linear map
$B_X: \operatorname{Sym}(n) \rightarrow \operatorname{Sym}(n)$ by
identifying $\operatorname{Sym}(n) $ with its dual using the inner product
$\langle\!\langle \cdot , \cdot \rangle\!\rangle$.

\begin{proposition}
\label{Poisson_tensors}
Denote the value at $X \in \operatorname{Sym}(n)$ of the Poisson tensors
corresponding to the  Lie-Poisson {\rm \eqref{LiePB}} and frozen {\rm
\eqref{FLiePB}} brackets by $B_X$ and $C_X$, respectively. Then for any $Y
\in \operatorname{Sym}(n)$ we have
\begin{align}
\label{LPtensor}
B_X(Y) & = XYN - NY X\\
\label{Ftensor}
C_X(Y) & = YN - NY.
\end{align}
\end{proposition}

\begin{proof}
Let $f$ and $g$ be locally defined smooth functions on
$\operatorname{Sym}(n)$. The definition of $B_X$ gives
\begin{align*}
\langle\!\langle\nabla f(X),B_X(\nabla g(X)\rangle\!\rangle&= \{f, g\}_N(X) 
\nonumber \\
& =  - \operatorname{trace}  \Big[ X \Big( \nabla f (X) N \nabla g (X) -
\nabla g (X) N \nabla f (X) \Big) \Big]  \\
& =   \operatorname{trace}  \Big[ \nabla f(X) \Big( X\nabla g (X) N  -
N \nabla g (X) X\Big) \Big]  \\
& = \langle\!\langle \nabla f(X), X\nabla g (X) N  -
N \nabla g (X) X \rangle\!\rangle ,
\end{align*}
which implies \eqref{LPtensor} since any $Y \in \operatorname{Sym}(n) $
is of the form $\nabla g(X)$, where $g(X) = \langle\!\langle X, Y
\rangle\!\rangle$. Similarly, the definition of $C_X $ gives
\begin{align*}
\langle\!\langle\nabla f(X),C_X(\nabla g(X)\rangle\!\rangle&= \{f,g\}_{FN}(X) 
\nonumber \\
& =  - \operatorname{trace} \Big( \nabla f (X) N \nabla g (X) -
\nabla g (X) N \nabla f (X) \Big)  \\
& =   \operatorname{trace}  \Big[ \nabla f(X) \Big(\nabla g (X) N  -
N \nabla g (X) \Big) \Big]  \\
& = \langle\!\langle \nabla f(X), \nabla g (X) N  -
N \nabla g (X) \rangle\!\rangle ,
\end{align*}
which proves \eqref{Ftensor}.
\end{proof}

\begin{proposition} 
\label{LPleaf}
Let $n = 2p+d$, where $2p = \operatorname{rank}N$.
The generic leaves of the Lie-Poisson bracket 
$\{ \cdot , \cdot \}_N$  are $2p(p+d)$-dimensional. 
\end{proposition}

\begin{proof}
As in the proof of Proposition \ref{SymStructure}, we orthogonally decompose $
\mathbb{R}^n = \operatorname{im} N \oplus \ker N $ so that $\bar{N} = N |
\operatorname{im} N : \operatorname{im} N \rightarrow \operatorname{im} N $ is an
isomorphism.  In this decomposition the matrix $N $ takes the form
\[
N = \begin{bmatrix}
\bar{N} & 0  \\
0 & 0  
 \end{bmatrix}
\]
and, according to the isomorphism $\Psi$ in Proposition \ref{SymStructure}, the
matrix $X $ can be written as 
\[
X = 
\begin{bmatrix}
S & A  \\
A^T & B  
 \end{bmatrix},
\]
where $S \in \operatorname{Sym}(2p)$, $B \in \operatorname{Sym}(d)$, and $A \in
\mathcal{M}_{(2p) \times d}$. Therefore, if 
\[
Y = 
\begin{bmatrix}
U & C  \\
C^T & D  
 \end{bmatrix} \in \operatorname{Sym}(n)
\]
with $U \in \operatorname{Sym}(2p)$, $D \in \operatorname{Sym}(d)$, $C \in
\mathcal{M}_{(2p) \times d}$, 
the Poisson tensor of the Lie-Poisson bracket $\{\cdot , \cdot \}_N$   takes the form
(see Proposition \ref{Poisson_tensors})
\begin{align*}
B_X(Y) &= XYN - NYX \\
&= 
\begin{bmatrix}
S & A  \\
A^T & B  
 \end{bmatrix}
 \begin{bmatrix}
U & C  \\
C^T & D  
 \end{bmatrix}
 \begin{bmatrix}
\bar{N} & 0  \\
0 & 0  
 \end{bmatrix} - 
 \begin{bmatrix}
\bar{N} & 0  \\
0 & 0  
 \end{bmatrix}
  \begin{bmatrix}
U & C  \\
C^T & D  
 \end{bmatrix}
 \begin{bmatrix}
S & A  \\
A^T & B  
 \end{bmatrix} \\
 & =  \begin{bmatrix}
S U \bar{N} - \bar{N} US + AC^T \bar{N} - \bar{N} CA^T 
& - \bar{N} U A- \bar{N} CB  \\
A^T U \bar{N} + BC^T \bar{N} & 0  
 \end{bmatrix}.
\end{align*}
Since $\bar{N}$ is invertible, the kernel of $B_X: \operatorname{Sym}(n) \rightarrow
\operatorname{Sym}(n)$ is therefore given by all $U \in \operatorname{Sym}(2p)$, $ C
\in \operatorname{Sym}(d)$, and $C \in \mathcal{M}_{(2p) \times d}$ such that 
\[
S U \bar{N} - \bar{N} US + AC^T \bar{N} - \bar{N} CA^T = 0
\quad \text{and} \quad 
UA  + CB = 0.
\]
To compute the dimension of the maximal symplectic leaves, we assume that  the matrix
$X $ is generic. So, supposing that $B $ is invertible, we have  $C = - UAB ^{-1}$
and
\[
\left(S - AB^{-1}A^T \right) U \bar{N} - \bar{N} U \left(S - AB^{-1}A^T \right) = 0.
\]

Since $S - AB^{-1}A^T \in \operatorname{Sym}(2p)$ is given, this condition is
identical to the vanishing of the Poisson tensor on the dual of the Lie algebra
$\left(\operatorname{Sym}(2p), [ \cdot \,, \cdot ]_{\bar{N}} \right)$ evaluated at 
$S - AB^{-1}A^T $. But $\bar{N}$ is invertible so, according to Proposition
\ref{sym_isomorphism}, this Lie algebra is isomorphic to $\mathfrak{sp}(2p,
\mathbb{R})$ whose rank is $p $. Therefore, the kernel of the map 
\[
U \in \operatorname{Sym}(2p) \mapsto
\left(S - AB^{-1}A^T \right) U \bar{N} - \bar{N} U \left(S - AB^{-1}A^T \right) \in
\operatorname{Sym}(2p)
\]
for generic $S - AB^{-1}A^T $ has dimension $p $.

Since $C = - UAB ^{-1}$ is uniquely determined and  $D \in \operatorname{Sym}(d)$ is
arbitrary, we see that the dimension of the kernel of $B_X$ for generic $X$ has
dimension $p +d(d+1)/2$.

Thus the dimension of the generic leaf of  the Lie-Poisson bracket $\{ \cdot , \cdot
\}_N$ is 
\[
\frac{1}{2}(2p+d)(2p+d+1) - p - \frac{1}{2}d(d+1) = 2p(p+d)
\]
as claimed in the statement of the proposition.
\end{proof}

\begin{proposition}
\label{frozen_leaves}
All leaves of the frozen Poisson bracket $\{ \cdot , \cdot \}_{FN}$ are 
\begin{itemize}
\item[{\rm (i)}] $2p(p+d)$-dimensional if $N $ is generic, that is, all its non-zero
eigenvalues are distinct, and 
\item[{\rm (ii)}] $p(p+1 + 2d)$-dimensional if all non-zero eigenvalue pairs  of $N $ are
equal.
\end{itemize}
\end{proposition}

\begin{proof}
Proceeding as in the proof of the previous proposition and using the same notation
for $N $, $X $, and $Y $, the Poisson tensor of the frozen bracket takes the form
 \begin{align*}
 C_X(Y) &= YN - NY
 =  \begin{bmatrix}
U & C  \\
C^T & D  
 \end{bmatrix}
 \begin{bmatrix}
\bar{N} & 0  \\
0 & 0  
 \end{bmatrix} - 
 \begin{bmatrix}
\bar{N} & 0  \\
0 & 0  
 \end{bmatrix}
  \begin{bmatrix}
U & C  \\
C^T & D  
 \end{bmatrix} \\
 & =   \begin{bmatrix}
U \bar{N} - \bar{N} U& \bar{N} C  \\
C^T \bar{N} & 0  
 \end{bmatrix}.
 \end{align*}
Thus, since $\bar{N}$ is invertible, the kernel of $C_X$ is given by all $U \in
\operatorname{Sym}(2p)$, $D \in \operatorname{Sym}(d)$, $C \in \mathcal{M}_{(2p)
\times d}$ such that $C = 0 $ and $U \bar{N} - \bar{N} U = 0 $. 

Since $\bar{N}$ is non-degenerate, there is an orthogonal matrix $Q$ such that 
\[
\bar{N} = Q^T 
\begin{bmatrix}
0 & V  \\
-V & 0  
 \end{bmatrix}
 Q,
 \]
 where  $V = \operatorname{diag}(v_1, \ldots , v_p)$  and $v _i\in  \mathbb{R}$, $v_i
\neq 0 $ for all $i = 1, \ldots , p$. Therefore,
\begin{align*}
 0 & = U \bar{N} - \bar{N} U = U Q^T 
 \begin{bmatrix}
0 & V  \\
-V & 0  
 \end{bmatrix} Q - Q^T
 \begin{bmatrix}
0 & V  \\
-V & 0  
 \end{bmatrix}
  Q U \\
 & = Q^T \left(Q U Q^T 
 \begin{bmatrix}
0 & V  \\
-V & 0  
 \end{bmatrix} - 
\begin{bmatrix}
0 & V  \\
-V & 0  
 \end{bmatrix} 
 QUQ ^T \right)Q
\end{align*}
 is equivalent to 
\begin{equation}
\label{FLPIntermediate}
\tilde{U}
 \begin{bmatrix}
0 & V  \\
-V & 0  
 \end{bmatrix} - 
 \begin{bmatrix}
0 & V  \\
-V & 0  
 \end{bmatrix}
 \tilde{U} = 0
\end{equation}
 where $\tilde{U} := QUQ^T \in \operatorname{Sym}(2p)$.
 Write 
 \[
\tilde{U} = 
 \begin{bmatrix}
U_{11} & U_{12}  \\
U_{12}^T & U_{22}  
 \end{bmatrix}
 \]
with $U_{11}$ and $U_{22}$ symmetric $p \times p $ matrices and $U_{12}$ an arbitrary
$p \times p $ matrix. Then 
\eqref{FLPIntermediate} is equivalent  to
\begin{equation}
\label{FLPIntermediate1}
U_{22} = VU_{11}V ^{-1} = V ^{-1} U_{11}V \quad \text{and} \quad U_{12}^T = - V ^{-1}
U_{12} V = - V U_{12} V ^{-1}.
\end{equation}

(i) Assume now that $v _i \neq v _j$ if $i \neq j $. Since $VU_{11}V ^{-1} = V ^{-1}
U_{11}V$ is equivalent to $V^2U_{11}V ^{-2} = U_{11}$, it follows that 
\[
\frac{v _i^2}{v _j^2} u_{11, ij} = u_{11, ij} \quad \text{for all} \quad i, j = 1,
\ldots , p,
\]
where $u_{11, ij}$ are the entries of the symmetric  
matrix $U_{11}$. Since the fraction on the left hand side is never equal to one for
$i \neq j$, this relation implies that 
$u_{11, ij} = 0$ for all $i \neq j$. Thus $U_{11}$ is diagonal and $U_{22} = U_{11}$.
A similar argument shows that $U_{12}$ is diagonal. However, then it follows that  
$U_{12} = -U_{12}^T$ which implies that $U_{12} = 0$.
Therefore, the kernel of the  map $U \mapsto U\bar{N} - \bar{N}U$ is $p
$-dimensional.

Concluding, the dimension of every leaf of the frozen Poisson structure equals
$\frac{1}{2}(2p+d)(2p+d+1) - p - \frac{1}{2}d(d+1) = 2p(p+d)$.

(ii) The other extreme case is when $v _i = v _j = : v $ for all $i,j = 1, \ldots, p
$. Then $V = v I $, where $I $ is the identity matrix, and \eqref{FLPIntermediate1}
becomes $U_{22} = U_{11}$, $U_{12}^T = -U_{12}$. Therefore, the kernel of the map $U
\mapsto U\bar{N} - \bar{N}U$ has dimension equal to
$\frac{1}{2}p(p+1) + \frac{1}{2}p(p-1) = p^2 $. 

Concluding, the dimension  of every leaf of the frozen Poisson structure equals
$\frac{1}{2}(2p+d)(2p+d+1) - p^2 - \frac{1}{2}d(d+1) = p(p+1+2d)$. 
\end{proof}

\begin{proposition} [Casimir Functions] Let the skew symmetric matrix $N$ have rank
$2p$  and size $n:= 2p+d$. Choose an orthonormal basis of $\mathbb{R}^{2p +d}$ in
which $N $ is written as
\[
N = 
\begin{bmatrix}
0 & V & 0  \\
-V & 0 & 0\\
0 & 0 & 0   
 \end{bmatrix},
 \]
 where $V$ is a real diagonal matrix whose entries are $v_1, \ldots, v_p$.

\begin{itemize}
\item[{\rm (i)}] If $v _i\neq v _j$ for all $i \neq j $, the $p + d(d+1)/2$ Casimir
functions for the frozen Poisson
structure \textup{\eqref{FLiePB}} are given by
\[
C^i _F (X) = \operatorname{trace} (E_i X), \quad i = 1, \ldots, p 
+ \frac{1}{2}d(d+1),
\]
where $E_i$ is any of the matrices
\[
\begin{bmatrix}
S_{kk} & 0 & 0 \\
0 & S_{kk} & 0 \\
0 & 0 & 0 
 \end{bmatrix}, \quad 
\begin{bmatrix}
0 & 0 & 0 \\
0 & 0 & 0 \\
0 & 0 & S_{ab} 
 \end{bmatrix}.
 \]
Here $S_{kk}$ is the $p \times p $ matrix all of whose  entries are zero except the
diagonal $(k,k)$ entry which is one and $S_{ab}$ is the $d \times d$ symmetric matrix
having  all entries equal to zero except for the $(a,b) $ and $(b,a)$ entries that
are equal to one.

\item[{\rm (ii)}] If $v _i = v _j$ for all $i,j=1, \ldots, p$, the $p^2 + d(d+1)/2$
Casimir functions for the frozen Poisson
structure \textup{\eqref{FLiePB}} are given by
\[
C^i _F (X) = \operatorname{trace} (E_i X), \quad i = 1, \ldots, p ^2 +
\frac{1}{2}d(d+1),
\]
where $E_i$ is any of the matrices
\[
\begin{bmatrix}
S_{kl} & 0 & 0 \\
0 & S_{kl} & 0 \\
0 & 0 & 0 
 \end{bmatrix}, \quad 
\begin{bmatrix}
0 & A_{kl} & 0 \\
-A_{kl} & 0 & 0 \\
0 & 0 & 0
 \end{bmatrix}, \quad 
\begin{bmatrix}
0 & 0 & 0 \\
0 & 0 & 0 \\
0 & 0 & S_{ab} 
 \end{bmatrix}.
\]
 Here $S_{kl}$ is the $p \times p$ symmetric matrix having  all entries equal to zero
except for the
$(k,l) $ and $(l,k)$ entries that are equal
to one and $A_{kl}$ is the $p \times p $ skew symmteric matrix with all entries
equal to zero except for the $(k,l)$ entry which is $1$ and the $(l,k)$ entry which
is $-1$.

\item[{\rm (iii)}] Denote 
\[
\bar{N} = \begin{bmatrix}
0 & V\\
-V & 0
\end{bmatrix} \quad \text{and} \quad
\hat{N}=\begin{bmatrix}
\bar{N}^{-1} & 0\\
0 & 0
\end{bmatrix}. 
\]
The $p + d(d+1)/2$ Casimir functions for the Lie-Poisson bracket  $\left\{ \cdot ,
\cdot \right\}_N$  are given by
\[
C^k(X) = \frac{1}{2k} \operatorname{trace} \left[ \left(X\hat{N}\right)^{2k} \right]
,
\quad
\text{for} \quad k = 1, \dots, p
\]
and
\[
C^k(X)=\operatorname{trace}(XE_k),
\quad\text{for} \quad k=p+1, \dots,p+\frac{1}{2}d(d+1)\,,
\]
where $E_k$ is any matrix of the form
\[
\begin{bmatrix}
0 & 0 & 0\\
0 & 0 & 0 \\
0 & 0 & S_{ab}
\end{bmatrix}.
\]
In the special case when $N$ is full rank the Casimirs are just
\[
C^k(X) = \frac{1}{2k} \operatorname{trace} \left[ \left(XN^{-1}\right)^{2k} \right]
,
\quad
\text{for} \quad k = 1, \dots, p,
\]
\end{itemize}
\end{proposition}

\begin{proof} To prove (i), recall from Proposition \ref{frozen_leaves}(i) that the
kernel of
the Poisson  tensor $C_X$ has dimension $p + \frac{1}{2}d(d+1)$. Moreover, if  $E$
belongs to this kernel, then the linear  function given by $X \mapsto
\operatorname{trace} (E X)$ has gradient $E$, which is annihilated by the
Poisson tensor $C_X$. Thus all $C^i_F$ are Casimir functions. Since the
gradients of all these functions are the $p + \frac{1}{2}d(d+1)$ matrices in
the statement which are obviously linearly independent, it follows that the functions
$C^i_F$
form a functionally independent set of Casimir functions for the frozen bracket
$\{ \cdot , \cdot \}_{FN}$. 

Part (ii) has an identical proof.

In the proof of (iii) we do not need the detailed $3 \times 3$ block decomposition of
$N $ and $X $ and shall use exclusively the $2 \times 2 $ block decomposition, where
the  $(1,1)$ block has size $(2p) \times (2p) $. Consider first the functions
$C^k(X)$ for $k=1,\dots, p$. Note that 
$\nabla C^k(X) = \hat{N} X \hat{N} \cdots \hat{N}X \hat{N}$
(with $(2k-1)$ factors of $X$) and hence \eqref{LPtensor} gives
\begin{equation}
B_X( \nabla C^k(X)) = X \left( \hat{N} X \hat{N} \cdots \hat{N} X \hat{N}
\right)N - N \left(\hat{N} X \hat{N} \cdots \hat{N} X \hat{N} \right)X.
\label{diff}
\end{equation}

Note firstly that in the case $N$ is invertible this is just
$ X \hat{N} \cdots \hat{N} X - X \hat{N}  \cdots \hat{N} X$
which is clearly $0$. 

Now consider the general case. We first observe that 
\begin{equation}
\hat{N}N=N\hat{N}=\begin{bmatrix}
I & 0\\
0 & 0
\end{bmatrix}.
\label{modinverse}
\end{equation}
The product of the last four factors in the first term of equation (\ref{diff}) is
thus
\[
\hat{N}X\hat{N}N=
\begin{bmatrix}
\bar{N}^{-1}S & 0\\
0 & 0
\end{bmatrix}.
\]
Similarly, the product of the first four factors of the second term
of (\ref{diff}) is
\[
N\hat{N}X\hat{N}=
\begin{bmatrix}
S\bar{N}^{-1} & 0\\
0 & 0
\end{bmatrix}.
\]
Continuing the multiplication in both terms in this fashion (always taking a group of
three consecutive factors from the  right and left, respectively) we see that 
both terms have only nonzero $(1,1)$ blocks which are identical and equal to
$S\bar{N}^{-1}S\bar{N}^{-1}....\bar{N}^{-1}S$. Thus, again, (\ref{diff}) is
identically zero. 

However,  $\mathfrak{sp}(2p, \mathbb{R})$  is identified with the subalgebra
consisting  of the $(1,1)$ blocks of elements of  $\operatorname{Sym}(n)$ (see
Proposition 
\ref{SymStructure}). The isomorphism $S \in \operatorname{Sym}(2p) \mapsto \bar{N} S
\in \mathfrak{sp}(2p, \mathbb{R}) $ given in Proposition \ref{sym_isomorphism}
identifies the basis of $p $ Casimirs in the dual of $\mathfrak{sp}(2p, \mathbb{R}) $
(given by the  even traces of the powers of a matrix) with the functions $S \mapsto
\operatorname{trace}\left[(S \bar{N}^{-1})^{2k} \right]/2k$. Therefore the functions
$C^k$ for $k = 1, \ldots, p$ given in the statement of the proposition are
functionally independent Casimirs for the Lie-Poisson bracket of
$\operatorname{Sym}(n)$.

To see that the remaining functions
$C^k(X)=\operatorname{trace}(XE_k)$ are Casimirs observe that  in this case
\[\nabla C^k(X)=\begin{bmatrix}
0 & 0 \\
0 & S_{ab}
\end{bmatrix}
\]
and
\begin{equation*}
B_X(\nabla C^k(X))=
\begin{bmatrix}
S & A\\
A^T & B
\end{bmatrix}
\begin{bmatrix}
0 & 0\\
0 & S_{ab}
\end{bmatrix}
\begin{bmatrix}
\bar{N} & 0\\
0 & 0
\end{bmatrix}
-
\begin{bmatrix}
\bar{N} & 0\\
0 & 0
\end{bmatrix}
\begin{bmatrix}
0 & 0\\
0 & S_{ab}
\end{bmatrix}
\begin{bmatrix}
S & A\\
A^T & B
\end{bmatrix}
=0.
\end{equation*}
Since the matrices $S_{ab}$ span the symmetric $k \times k $ matrices, these Casimirs
are functionally independent. The two sets of Casimirs are also independent taken
together, since  each set depends only on a subset of independent variables and these
two sets of variables are disjoint. We have thus obtained $p + d(d+1)/2$ Casimirs,
which is the  codimension of the generic leaf thus proving  that they generate the
space of all Casimir functions of the Lie-Poisson bracket.
 \end{proof}

\paragraph{The equations in the degenerate case.} If $N $ is degenerate, representing
 it and the matrix $X \in \operatorname{Sym}(n)$ as in Proposition
\ref{SymStructure}, the equations $\dot{X} = [X^2, N]$ are equivalent  to the system
\[
\left\{
\begin{aligned}
\dot{S} &= [S^2 + A^T A, \bar{N}]\\ 
\dot{A} & = - \bar{N}(SA + A B)\\
\dot{B} & = 0
\end{aligned}
\right.
\]

\section{The Sectional Operator Equations}
\label{Sec}
This section shows that  the flow \eqref{eq:1.1} is not of the sectional
operator type discussed in \cite{MiFo1976}; in fact, this is the
case already for $2 \times 2 $ matrices with the canonical choice of $N$.

Let
\begin{equation}
N=\left[\begin{array}{cc}
0 & 1\\
-1& 0
\end{array}
\right]
\end{equation}
and denote elements of $\operatorname{Sym} (2)$ by 
\begin{equation}
X=
\begin{bmatrix}
a & b\\
b& d
\end{bmatrix}, \quad a, b, c \in \mathbb{R}.
\end{equation}

One can readily check that a maximal Abelian subalgebra of
$\operatorname{Sym}(2)$, that is, a Cartan subalgebra,  consists of purely
off diagonal matrices 
\begin{equation}
A=
\begin{bmatrix}
0 & \alpha\\
\alpha& 0
\end{bmatrix}, \quad \alpha\in \mathbb{R}.
\end{equation}
A complementary subspace is $\operatorname{Sym}_d (2)$, the space of diagonal $2
\times 2 $ matrices. 
Notice that for any $X \in \operatorname{Sym}(2) $ we have
\begin{equation}
[A,X]_N=
\begin{bmatrix}
-2\alpha a& 0\\
0& 2\alpha d
\end{bmatrix}
\end{equation}
and hence, also in accordance with general theory, if $\alpha \neq 0$, then
$\operatorname{ad}_A :  \operatorname{Sym}_d (2) \rightarrow
\operatorname{Sym}_d (2) $ is an isomorphism. Thus the
inverse $\operatorname{ad}_A ^{-1}: \operatorname{Sym}_d(2) \rightarrow
\operatorname{Sym}_d(2)$ is defined and hence 
\begin{equation}
\operatorname{ad}_A^{-1}\left(\operatorname{ad}_B X\right)
=
\frac{\beta}{\alpha}
\begin{bmatrix}
a & 0\\
0& d
\end{bmatrix} \quad \text{for} \quad 
A = \begin{bmatrix} 0 & \alpha \\ \alpha & 0 \end{bmatrix}, \quad 
B = \begin{bmatrix} 0 & \beta \\ \beta & 0 \end{bmatrix}, \quad \alpha \neq 0.
\end{equation}
An operator of this form is called a \emph{sectional operator} in the sense of
\cite{MiFo1976}. The equations defined by a sectional operator are
 \begin{equation}
\label{MSeq}
 \dot{X} = \left[ X, \operatorname{ad}_A^{-1}\left(\operatorname{ad}_B X\right)
\right] _N
 =  \frac{\beta }{ \alpha } \begin{bmatrix}
 - 2 ab & 0 \\
 0 & 2 bd 
 \end{bmatrix}.
 \end{equation}

\textit{We shall now prove that} \eqref{eq:1.1} \textit{is not in this
family.} Indeed, since 
\begin{equation}
XN+NX=
\begin{bmatrix}
0 & a+d\\
-a-d& 0
\end{bmatrix}
= (a+d)N
\end{equation}
equation \eqref{eq:1.2} becomes
\begin{equation}
\label{oureq}
\dot{X} = \left[ X , XN + NX \right]   =  
(a + d) \begin{bmatrix} - 2 b & a - d \\ a - d & 2b \end{bmatrix}.
\end{equation}
The only way equations \eqref{MSeq} and \eqref{oureq} can be identical is if
one requires that $a = d$, which is not allowed since $X $  is arbitrary in
$\operatorname{Sym}(2) $. Therefore the system \eqref{eq:1.1} \textit{is not
in the list of equations of generalized rigid body type on
$\mathfrak{sp}(2, \mathbb{R})$ described by a sectional operator in}
\cite{MiFo1976}.

Despite the fact that our system is not in the class of integrable systems 
studied in \cite{MiFo1976}, we shall see in the next sections that by using
the techniques of \cite{Manakov1976} and \cite{Magri1978} (the method of
recursion operators), the system is nonetheless integrable.

\section{Lax Pairs with Parameter}
\label{Lax}
To prove that system \eqref{eq:1.1} is integrable for any choice of $N$, we
will compute its flow invariants. Bear it in mind that, by virtue of the
isospectral representation \eqref{eq:1.2}, we already know that the
eigenvalues of $X$, or alternatively, the quantities $\operatorname{trace}
X^k$ for $k=1,2,\ldots,n-1$, are invariants.

One way to compute additional invariants is to rewrite the system as a
Lax pair with a parameter. One can do this in a fashion  similar
to that  for the generalized rigid body equations (see
\cite{Manakov1976}). 

\begin{theorem}
  \label{th:2}
  Let $\lambda$ be a real parameter. The system {\rm \eqref{eq:1.2}} is
equivalent to the following Lax pair system  
  \begin{equation}
    \label{eq:3.1}
 \frac{d}{dt}   (X+\lambda N)  = \left[ X+\lambda N, NX+XN+\lambda N^2 \right]
   \end{equation}
\end{theorem}

 \begin{proof}
  The proof is a computation. The only nontrivial power of $\lambda$ to check is the
first
one. In fact,    the coefficient of $\lambda$ on the right hand side of 
  equation \eqref{eq:3.1} is  
\begin{align*}
     &  [N, NX+XN]+[X, N^2] \\
 &  \qquad \qquad  =  N ^2 X +N XN -NXN - X N ^2 + X N ^2 - N ^2 X   = 0,
\end{align*}
which proves \eqref{eq:3.1}.
\end{proof}

We recall from \cite{Manakov1976} and \cite{Ratiu1980} that the
left-invariant generalized rigid body equations on $SO(n)$ may be
written as 
\begin{equation}
  \label{eq:3.2}
\dot{  M} = [M,\Omega],  \qquad M(0)=M_0\in\mathfrak{so}(n),
\end{equation}
where $\Omega=Q^{-1} \dot{Q} \in \mathfrak{so}(n)$ is the body
angular velocity, $Q\in SO(n)$ denotes the configuration space variable (the
attitude of the body),  and 
\begin{displaymath}
  M=J(\Omega) :=\Lambda\Omega +\Omega\Lambda \in \mathfrak{so}(n)
\end{displaymath}
is the body angular momentum. Here $J: \mathfrak{so}(n) \rightarrow 
\mathfrak{so}(n)
$ is
the symmetric, positive definite (and hence invertible) operator
defined by 
\begin{displaymath}
  J(\Omega):=\Lambda\Omega +\Omega\Lambda ,
\end{displaymath}
where $\Lambda$ is a diagonal matrix satisfying $\Lambda_i + \Lambda_j
>0$ for all $i \neq j$. For $n=3$ the elements of $\Lambda_i$
are related to the standard diagonal moment of inertia tensor $I$ by
$I_1 = \Lambda_2 + \Lambda_3$,  $I_2 = \Lambda_3 + \Lambda_1$,
$I_3 = \Lambda_1 + \Lambda_2$.

\cite{Manakov1976} has noticed that the generalized rigid body equations 
\eqref{eq:3.2} can be written as a Lax equation
with a parameter in the form
\begin{equation} 
  \label{eq:3.3} 
  \frac{d}{dt}(M+\lambda \Lambda^2)= [M+\lambda
  \Lambda^2,\Omega+\lambda \Lambda]. 
\end{equation} 

Note the following contrast with our setting: in the Manakov case
the system matrix $M$ is in $\mathfrak{so}(n)$
and the parameter $\Lambda$ is a symmetric matrix while in our case $X$ is
symmetric and the parameter $N\in\mathfrak{so}(n)$. 

For the generalized rigid body the nontrivial coefficients of
$\lambda ^i, 0<i<k$ in the traces of the powers of $M+\lambda \Lambda^2$ then
yield the right number of independent integrals in involution to prove
integrability of the flow on a generic adjoint orbit of $SO(n)$
(identified with the corresponding coadjoint orbit). The case $i = 0 $ needs
to be eliminated, because these are Casimir functions.

Similarly, in our case, the nontrivial coefficients of $\lambda ^i, 0\leq
i\leq k,$ in
\begin{equation}
\label{consquant}
h_k^\lambda(X):= \frac{1}{k}  \operatorname{trace} (X+\lambda N)^k,\qquad
k=1,2,\ldots,n-1
\end{equation}
yield the conserved quantities. The coefficient of
$\lambda ^r, 0\leq r\leq k$, in
\eqref{consquant} is
\begin{displaymath}
  \operatorname{trace} \sum_{|{i}|=k - r} \sum_{|{j}|=r} X^{i_1} N^{j_1}
X^{i_2}
  \cdots X^{i_s} N^{j_s} , \qquad r=0,\ldots,k,\quad
  k=1,\ldots,n-1,
\end{displaymath}
where ${i} = (i _1, i _2, \ldots i _s)$, $j = (j_1, j_2, \ldots j_s)$ are
multi-indices,  $i _q, j _q= 0, 1, \ldots, k$, and
$|{i}|=\sum_{q = 1 } ^s  i_q$, $|j| =\sum_{q = 1 } ^s  j_q$. The
coefficient of $\lambda^k$ is the constant $N^k$ so it should  not be
counted. Thus we have $r < k $. In addition, since the trace of a matrix
equals the trace of its transpose,
$X\in\operatorname{Sym}(n)$, and $N\in\mathfrak{so}(n)$, it follows that
\begin{displaymath}
  \operatorname{trace}  X^{i_1} N^{j_1} X^{i_2} \cdots X^{i_s}  N^{j_s}=
  (-1)^{|{j}|} \operatorname{trace} N^{j_s} X^{j_s} \cdots X^{i_2} N^{j_1} X^{i_1}. 
\end{displaymath}
Therefore, if $r$ is odd, then necessarily 
\begin{displaymath}
  \operatorname{trace} \sum_{|{i}|=k-r} \sum_{|{j}|=r} X^{i_1} N^{j_1}
X^{i_2}
  \cdots X^{i_s} N^{j_s}=0
\end{displaymath}
and only for even $r$ we get an invariant. Thus, we are left with
the invariants
\begin{equation}
  \label{eq:3.4}
h_{k,2r}(X): =   \operatorname{trace} \sum_{|{i}|=k-2r} \sum_{|{j}|=2r} X^{i_1}
N^{j_1} X^{i_2}
  \cdots X^{i_s} N^{j_s}
\end{equation}
for $k = 1, \dots, n-1$, $i_q = 1, \ldots, k$, $j_q = 0, \ldots, k-1$, $r=0,\ldots,
  \left[\frac{k-1}{2} \right]$, where $[\ell]$ denotes the
integer part of $\ell \in \mathbb{R}$. 

The integrals \eqref{eq:3.4} are thus the coefficients of $\lambda^{2r}$, $0 < 2r <
k$, in the expansion of $\frac{1}{k}\operatorname{trace}(X + \lambda N)^k$. For
example, if $k = 1$ or $k=2$ then we have one  integral, the ceofficient of
$\lambda^0$. If $k = 3$ or $k=4$, only the coefficients of $\lambda^2$ and
$\lambda^0$ yield non-trivial integrals. If $k=5$ or $k=6$ it is the coefficients of
$\lambda^4$, $\lambda^2$, and $\lambda^0$ that give non-trivial integrals. In
general, for the power $k $,  we have $\left[\frac{k+1}{2}\right]$ integrals. Recall
that $k = 1, \ldots, n-1$. If $n - 1 = 2 \ell$, we have  hence 
\begin{align*}
&1+1 + 2 + 2 + \dots + \left[\frac{n-1+1}{2} \right] + \left[\frac{n-1+1}{2} \right]
=
1+1 + 2 + 2 + \dots + \ell + \ell \\
& \qquad = \ell(\ell + 1) = \frac{n-1}{2} \left(\frac{n-1}{2} + 1 \right) = 
\frac{n-1}{2}  \frac{n+1}{2}
\end{align*}
integrals. If  $n-1 = 2 \ell +1$ then we have
\begin{align*}
&1+1 + 2 + 2 + \dots + \left[\frac{n-2+1}{2} \right] + \left[\frac{n-2+1}{2} \right] 
+ \left[\frac{n-1+1}{2} \right] \\
& \qquad =
1+1 + 2 + 2 + \dots + \ell + \ell + (\ell + 1) \\
& \qquad = \ell(\ell + 1) + (\ell + 1) = (\ell + 1)^2 = \left(\frac{n}{2} \right)^2
\end{align*}
integrals. However, 
\[
\left[\phantom{\frac{n+1}{2}} \hspace{-.37in}
\frac{n}{2}\right] \left[\frac{n+1}{2} \right] = \left\{
\begin{aligned}
\frac{n-1}{2}  \frac{n+1}{2}, \quad  \text{if} \quad  n \quad \text{is odd}\\
\left(\frac{n}{2} \right)^2, \quad \; \text{if} \quad n \quad \text{is even}
\end{aligned}
\right.
\]
Concluding we have 
\[
\left[\phantom{\frac{n+1}{2}} \hspace{-.37in}
\frac{n}{2}\right]\left[\frac{n+1}{2}\right] 
\]
invariants which are  the coefficients of $\lambda^{2r}$, $0 < 2r < k$, in the
expansion of $\frac{1}{k}\operatorname{trace}(X + \lambda N)^k$ for $k = 1, \ldots ,
n-1$.

Are these integrals the right candidates to prove complete integrability of the
system $\dot{X} = [X^2, N]$?
\begin{itemize}
\item If $N $ is invertible, then $n = 2p$ and hence
\begin{align*}
\left[\phantom{\frac{n+1}{2}} \hspace{-.38in}
\frac{n}{2}\right]\left[\frac{n+1}{2}\right] 
&= \left[\frac{2p}{2} \right] \left[\frac{2p+1}{2} \right] = p^2 =
\frac{1}{2} \left( 2p^2 + p - p \right)\\
&= \frac{1}{2} \left(\dim \mathfrak{sp}(2p, \mathbb{R}) -
\operatorname{rank}\mathfrak{sp}(2p,\mathbb{R}) \right)
\end{align*}
which is half the dimension of the generic adjoint orbit in $\mathfrak{sp}(2p,
\mathbb{R})$. Therefore, these conserved quantities are the right candidates
to prove that this system is integrable on the generic coadjoint orbit of
$\operatorname{Sym}(n)$. This will be proved in the next sections.

\item If $N $ is non-invertible (which is equivalent  to $d \neq  0$), then $n = 2p +
d$ and hence
\begin{align*}
\left[\phantom{\frac{n+1}{2}} \hspace{-.38in}
\frac{n}{2}\right]\left[\frac{n+1}{2}\right] 
&= \left[\frac{2p+d}{2} \right] \left[\frac{2p+d+1}{2} \right] \\
& = \left(p + \left[\phantom{\frac{n+1}{2}} \hspace{-.38in}
\frac{d}{2}\right] \right) \left(p + \left[\frac{d+1}{2}\right]  \right)\\
&= p^2 + p \left(\left[\phantom{\frac{n+1}{2}} \hspace{-.38in}
\frac{d}{2}\right] +  \left[\frac{d+1}{2}\right] \right) + 
\left[\phantom{\frac{n+1}{2}} \hspace{-.38in}
\frac{d}{2}\right]  
\left[\frac{d+1}{2}\right]\\
& = p^2 + pd + \left[\phantom{\frac{n+1}{2}} \hspace{-.38in}
\frac{d}{2}\right]  
\left[\frac{d+1}{2}\right].
\end{align*}
The right number of  integrals is $p(p+d)$ according to Proposition \ref{LPleaf}, so
this calculation seems to indicate that there are additional integrals. The situation
is not so simple since there are redundancies due to the degeneracy of $N $. Note,
however, that if $d=1$, then  we do get the right number of integrals. We shall
return to the study of the degenerate case in \S\ref{sec:independence}. 
\end{itemize}

\section{Involution}
\label{Involution}

In this section we prove involution of the integrals found in the previous section
for arbitrary $N \in \mathfrak{so}(n)$.

\paragraph{Bi-Hamiltonian structure.} We begin with the following observation.

\begin{proposition}
The system $\dot{X} = X ^2 N - N X ^2$ is Hamiltonian with respect to the
bracket $\left\{ f, g \right\} _N $ defined in {\rm \eqref{LiePB}} using the
Hamiltonian  $h _2 (X) := \frac{1}{2} \operatorname{trace} (X ^2) $ and is
also Hamiltonian with respect to the compatible bracket $\left\{ f, g \right\}
_{FN} $ defined in {\rm \eqref{FLiePB}} using the Hamiltonian $h _3 (X) :=
\frac{1 }{3 } \operatorname{trace} (X ^3) $. 
 \end{proposition}

\begin{proof} We have implicitly checked the first statement already using 
Euler-Poincar\'e theory, but here is a direct verification. We want to
show that the condition $ \dot{f} = \left\{ f, h _2 \right\}_N$ for any $f$
determines the equations $\dot{X} = X ^2 N - N X ^2$. First note that
$\dot{f} = \operatorname{trace} ( \nabla f (X) \dot{X } ) $. Second, since
$\nabla h _2 (X) = X $, the right hand side $ \left\{ f, h _2 \right\}_N$
becomes by \eqref{LiePB}
\begin{align*}
 \left\{ f, h _2 \right\}_N(X) 
& = - \operatorname{trace} \Big[ X \Big( \nabla f(X) N X - X N \nabla f(X)
\Big) \Big] \\
& = - \operatorname{trace}     \Big( \nabla f(X) N X ^2 -   \nabla f(X) X ^2
N\Big) . 
\end{align*}
Thus, $ \dot{X} = X ^2 N - N X ^2$ as required.

To show that the same system is Hamiltonian in the frozen structure, we 
proceed in a similar way. Noting that $\nabla h _3 (X) = X ^2 $, we get from
\eqref{FLiePB}
\begin{align*}
 \left\{ f, h _3 \right\}_{FN} (X) & = - \operatorname{trace}  \Big( \nabla f N X^2
-
X^2 N \nabla f \Big)   \\
 & = - \operatorname{trace}     \Big( \nabla f N X ^2 -   \nabla f X ^2 N \Big) , 
\end{align*}
and hence $ \dot{X} = X ^2 N - N X ^2$, as before.
\end{proof}

\paragraph{Involution.} Next we begin the proof that the
$\left[\frac{n}{2}\right]\left[\frac{n+1}{2}\right] $ integrals given in
\eqref{eq:3.4}, namely
\[
h_{k,2r}(X) :=  \operatorname{trace} \sum_{|{i}|=k-2r} \sum_{|{j}|=2r}
X^{i_1} N^{j_1} X^{i_2}
  \cdots X^{i_s} N^{j_s},
\]
where  $k = 1, \dots, n-1$, $i_q = 1, \ldots, k$, $j_q = 0, \ldots, k-1$,
$r=0,\ldots,
  \left[\frac{k-1}{2} \right]$, are in involution. It will be convenient below to
write the expansion of $h_k ^\lambda$ starting with the highest power of $\lambda$,
that is,
\begin{equation}
h_k^{\lambda}(X)=
\frac{1 }{ k } \operatorname{trace} \left( X + \lambda N \right) ^k
=\sum_{r=0}^k\lambda^{k-r}h_{k,k-r}(X)\,.
\label{lambdaexpansion}
\end{equation}
As explained before, not all of these coefficients should be counted: roughly
half of them vanish and the last one, namely, $h_{k,k}$, is the constant
$N^k$. Consistent with our notation for the Hamiltonians, we set $h_k =
h_{k,0} $.

Firstly we need the gradients of the functions $h^{\lambda}_k$.

\begin{lemma}
The gradients  $\nabla h^{\lambda}_k$ are given by
\begin{equation}
\nabla h^{\lambda}_k(X)
=\frac{1}{2}(X+\lambda N)^{k-1}+\frac{1}{2}(X-\lambda N)^{k-1}.
\end{equation}
\end{lemma}
\begin{proof}
We have for any $Y \in \operatorname{Sym}(n)$,
\begin{align*}
\langle\!\langle \nabla h^{\lambda}_k(X), Y \rangle\!\rangle&=
\mathbf{d} h^{\lambda}_k(X) \cdot Y = 
\operatorname{trace}\left((X+\lambda N)^{k-1}Y\right) \\
&=\frac{1}{2}\operatorname{trace}
\left(\left((X+\lambda N)^{k-1}+(X-\lambda N)^{k-1}\right)Y\right).
\end{align*}
Since $\langle\!\langle \,, \rangle\!\rangle$ is
nondegenerate on $ \operatorname{Sym}(n)$, the result follows. 
\end{proof}

\begin{proposition}
\label{lambdarecursionproposition}
\begin{equation}
B_X(\nabla h^{\lambda}_k(X))=C_X(\nabla h^{\lambda}_{k+1}(X))
\label{lambdarecursion}
\end{equation}
\end{proposition}
\begin{proof}
By \eqref{LPtensor} we have
\begin{align*}
&B_X(\nabla h^{\lambda}_k(X))= 
X \nabla h^{\lambda}_k(X) N - N \nabla h^{\lambda}_k(X) X  \\
& \quad = \frac{1}{2}\left[X(X+\lambda N)^{k-1}N + X(X-\lambda N)^{k-1}N
\right. \\ 
& \qquad \qquad \quad \left.
- N(X+\lambda N)^{k-1}X - N(X-\lambda N)^{k-1}X\right] \\
&  \quad 
= \frac{1}{2}\left[(X+\lambda N)^{k}N - \lambda N (X+\lambda N)^{k-1}N
+ (X-\lambda N)^{k}N + \lambda N (X-\lambda N)^{k-1}N \right. \\
& \quad \qquad \left. 
- N(X+\lambda N)^{k} + \lambda N (X+\lambda N)^{k-1}N
- N(X-\lambda N)^{k} - \lambda N (X-\lambda N)^{k-1}N \right] \\
& \quad = \frac{1}{2}\left[(X+\lambda N)^{k}N + (X-\lambda N)^{k}N -
N(X+\lambda N)^{k} - N(X-\lambda N)^{k}\right] \\
& \quad = \nabla h^{\lambda}_{k+1}(X) N - N \nabla h^{\lambda}_{k+1}(X) =
C_X(\nabla h^{\lambda}_{k+1}(X))
\end{align*}
by \eqref{Ftensor}, which proves the formula.
\end{proof}

\begin{proposition}
\label{recursionproposition}
The functions $h_{k,k-r}$ satisfy the recursion relation
\begin{equation}
B_X(\nabla h_{k,k-r}(X))=C_X( \nabla h_{k+1,k-r}(X))
\label{recursion}
\end{equation}
\end{proposition}
\begin{proof}
Substituting (\ref{lambdaexpansion}) into (\ref{lambdarecursion})
we obtain
\[
\sum_{r=0}^k\lambda^{k-r}B_X\left(\nabla h_{k,k-r}(X)\right)
=\sum_{r=0}^{k+1}\lambda^{k+1-r}C_X\left(\nabla h_{k+1,k+1-r}(X)\right).
\]
Since $\nabla h_{k+1, k+1}(X) = N^{k+1}$, formula \eqref{Ftensor} implies
that $C_X\left(\nabla h_{k+1,k+1}(X)\right) = 0 $. Thus on the right hand
side the sum begins at $r = 1 $. Changing the summation index on the right 
hand side now from $r $ to $r-1 $ and identifying the coefficients of like
powers of $\lambda$ yields \eqref{recursion}.
\end{proof}

\noindent {\bf Remark.}
It is worth making a few remarks about Propositions
\ref{lambdarecursionproposition} and \ref{recursionproposition}.
Note that unlike the similar recursion for the rigid body Mankov
integrals (see e.g. \cite{MoPi1996}) our polynomial recursion relation 
(\ref{lambdarecursion}) does not have a premultiplier $\lambda$ on the 
right hand side and the polynomials on the left and right hand sides
appear to be of different order. This cannot be and indeed is not so. 
Indeed the highest order order coefficient on the right hand side
vanishes by virtue of following result.

\begin{corollary}
The functions $h_{k,k-1}(X)$ are Casimirs for the frozen Poisson
structure, i.e.
\begin{equation}
\label{Fcasimir}
C_X\left(\nabla h_{k,k-1}(X)\right)=0
\end{equation}
for all $k$.
\end{corollary}

\begin{proof} By \eqref{lambdaexpansion}, $h_{k,k-1}(X) =
\operatorname{trace} \left(N^{k-1} X \right)$, so its gradient equals  
$\nabla h_{k,k-1}(X) = N^{k-1} $. So \eqref{Ftensor} immediately gives
\eqref{Fcasimir}.
\end{proof}

The recursion relations \eqref{recursion} for $r=0$ also imply the following
relation between the Hamiltonians that can also be easily checked  by hand.

\begin{corollary}
\begin{equation}
B_X\left(\nabla h_{k}(X)\right)
=C_X\left(\nabla h_{k+1}(X)\right)
\end{equation}
\end{corollary}

\noindent {\bf Example:} An interesting nontrivial example of the recursion
relation to check is
$B_X(dh_{3,2}(X))=C_X(dh_{4,2}(X))$ where
$h_{3,2}(X)=\operatorname{trace}(N^2X)$
and
$h_{4,2}(X)=\operatorname{trace}(N^2X^2)+\frac{1}{2}\operatorname{trace}(NXNX)$.
This
example illustrates how the recursion relation works despite
the apparent inconsistency in order.

\bigskip

Uising the recursion relations involution follows immediately.
 \begin{proposition} The invariants $h _{k,k - r} $ are in involution with 
respect to both Poisson brackets $\left\{f, g \right\}_N$ and $\left\{f,
g \right\}_{FN} $.
 \end{proposition}
\begin{proof}
The definition of the Poisson tensors $B_X $ and $C_X $ and the recursion
relation \eqref{recursion} give
\begin{align*}
\left\{h_{k, k-r},h_{l, l-q}\right\}_N
& = \langle\!\langle\nabla h_{k, k-r}(X),B_X(\nabla h_{l,l-q}(X))
\rangle\!\rangle   \\
& = \langle\!\langle\nabla h_{k, k-r} (X),C_X(\nabla
h_{l+1, l-q}(X))\rangle\!\rangle  \\ 
&= \left\{h_{k, k-r} ,h_{l+1, l-q}\right\}_{FN}
=-\left\{h_{l+1, l-q},h_{k, k-r}\right\}_{FN} \\
&= - \langle\!\langle\nabla h_{l+1, l-q}(X),C_X(\nabla
h_{k, k-r}(X))\rangle\!\rangle \\ 
& = - \langle\!\langle\nabla h_{l+1, l-q}(X),B_X(\nabla h_{k-1, k-r}(X))
\rangle\!\rangle \\
& =-\left\{h_{l+1, l-q},h_{k-1, k-r}\right\}_N 
=\left\{h_{k-1, k-r},h_{l+1, l-q}\right\}_N
\end{align*}
for any $k,l= 1, \dots, n-1$, $r = 1, \dots, k$ and $q = 0, \dots, l-1$. Of
course, in these relations we assume that $k-r$ and $l-q$ are even, for if
at least one of them is odd, the identity above has zeros on both sides. 
Repeated application of this relation eventually leads to Hamiltonians $h_{k,
k-r} $ where either $k-r$ is a power of $\lambda$ that does not exist for $k$, in
which
case the Hamiltonian  is zero, or one is led to $h_{0, 0} $ which is
constant. This shows that $\left\{h_{k, k-r},h_{l, l-q}\right\}_N = 0 $ for
any pair of indices. 

In a similar way one shows that $\left\{h_{k, k-r},h_{l, l-q}\right\}_{FN} =
0 $.
\end{proof}

\section{Independence}
\label{sec:independence}

To complete the proof of integrability we need to show that the integrals
 $h_{k,2r}$ are independent. We will demonstrate this first in the generic
case that $N$ is invertible with distinct eigenvalues. 
 
By \eqref{eq:3.4}, the gradients of the integrals $h_{k,2r}$ have the expression
\begin{equation}
  \label{gradients_of_invariants}
\nabla h_{k,2r}(X): =   \sum_{|{i}|=k-2r-1} \sum_{|{j}|=2r} X^{i_1} N^{j_1} X^{i_2}
  \cdots X^{i_s} N^{j_s}
\end{equation}
where  $k = 1, \dots, n-1$, $i_q = 1, \ldots, k$, $j_q = 0, \ldots, k-1$,
$r=0,\ldots,
  \left[\frac{k-1}{2} \right]$.

\paragraph{The Generic Case.} We consider the case $N$ invertible with distinct
eigenvalues. Therefore $d= 0$ and $n=2p$. In this case we show that the integrals
$h_{k,2r}$ given in (\ref{eq:3.4}) are independent, and hence the
system (\ref{eq:1.1})  is system is integrable. 

\begin{theorem}
For generic $N$ the integrals  $h_{k,2r}$ given by equation
(\ref{eq:3.4}) are independent.
\end{theorem}

\begin{proof}
We are concerned with the linear independence (in a generic sense) of
(\ref{gradients_of_invariants})
where $k=1,\ldots,n-1$, $i_q=1,\ldots,k$, $j_q=0,\ldots,k-1$ and
$r=0,\ldots [\frac12(k-1)]$. We assume that $N$ is invertible with
distinct eigenvalues and, without loss of generality, that $X$ is
diagonal, 
\begin{displaymath}
  X=\mathrm{diag}\,\mu.
\end{displaymath}

This reduces the problem to a problem 
about the independence of polynomials in single matrix variable. 

Now, we aim to prove a stronger statement: the terms
\begin{displaymath}
  v_{i,j}= X^{i_1} N^{j_1} X^{i_2}\cdots X^{i_s} N^{j_s}
\end{displaymath}
are independent for all multi-indices $i$ and $j$ in the above
range. Note however that each $v_{i,j}$ is a $q$-degree polynomial in
$\mu_1,\mu_2,\ldots,\mu_n$, where $q=k-2r-1\in\{0,\ldots,n-2\}$. Let
\begin{displaymath}
  \mathcal{H}_q=\{ v_{i,j}\,:\, |i|=q,\; |j|\,\mbox{even} \}. 
\end{displaymath}
Clearly, in a generic sense, if linear dependence exists, it must exist
\textit{within\/} a set $\mathcal{H}_q$. In other words, if we can
prove that there is no linear dependence within each $\mathcal{H}_q$,
we are done. 
(Note that since $k\le n-1$ in the expression (\ref{gradients_of_invariants})
there is no dependence of powers of $X$ on lower powers through 
the characteristic polynomial.)

There is nothing to prove for $q=0$ For $q=1$ we have 
\begin{displaymath}
  \mathcal{H}_1=\{XN^j\,:\, j\mbox{\ even}\} \cup \{N^jX\,:\, j\mbox{\
    even}\}. 
\end{displaymath}
Suppose that there exists linear dependence in $\mathcal{H}_1$. Then
there necessarily exist $\rho_0,\rho_2,\ldots,\rho_{n-2}$ and
$\kappa_0,\kappa_2,\ldots, \kappa_{n-2}$, not all zero, such that 
\begin{displaymath}
  X\left(\sum \rho_{2j} N^{2j}\right) +\left(\sum \kappa_{2j}
  N^{2j}\right) X=0 =XR(N)+K(N)X=0.
\end{displaymath}
Therefore,
\begin{displaymath}
  \mu_a [R(N)]_{a,b} +[K(N)]_{a,b} \mu_b=0,\qquad a,b=1,\ldots,n.
\end{displaymath}
Generically (i.e., for all $\mu$ except for a set of measure zero)
this can hold only if $R(N),K(N)=0$. But $\deg R,\deg K\leq n-1$ and,
since the eigenvalues of $N$ are distinct, the degree of the minimal
polynomial of $N$ is $n$. Therefore $K,R\equiv0$, a
contradiction. Hence there is no linear dependence.

We continue to $s=2$. Now
\begin{displaymath}
  \mathcal{H}_2=\{ X^{i_1} N^{j_1} X^{i_2} N^{j_2} X^{i_3}\,:\,
  i_1+i_2+i_3=2,\; j_1+j_2\mbox{\ even}\}.
\end{displaymath}
Assume that there exist $\rho_{i,j}$, not all zero, s.t.
\begin{displaymath}
  \sum_{i,j} \rho_{i,j} X^{i_1} N^{j_1} X^{i_2} N^{j_2} X^{i_3}=0.
\end{displaymath}
Therefore
\begin{displaymath}
  \sum_{i,j} \rho_{i,j} \sum_b \mu_a^{i_1} \mu_b^{i_2} \mu_c^{i_3}
  (N^{j_1})_{a,b} (N^{j_2})_{b,c}=0,\qquad a,c=1,\ldots,n. 
\end{displaymath}
Note that we want the above to hold for all real $\mu_k$, but this is
possible only if 
\begin{displaymath}
  0=\sum_{i,j} \rho_{i,j} \sum_b (N^{j_1})_{a,b} (N^{j_2})_{b,c}= \sum_{i,j}
  \rho_{i,j} (N^{j_1+j_2})_{a,c},\qquad a,c=1,\ldots,n,
\end{displaymath}
thus
\begin{displaymath}
  \sum_{i.j} \rho_{i,j} N^{j_1+j_2}=0.
\end{displaymath}
We again obtain a polynomial in $N^2$ of degree $< n/2$, which cannot
be zero: a contradiction.

We can continue for higher $s$ in an identical manner.
\end{proof}

Hence, since we have involution and independence we 
have proved the following.
\begin{theorem}
For $N$ invertible with distinct eigenvalues the system \textup{(\ref{eq:1.1})} is completely integrable.
\end{theorem}

\begin{corollary}
For $N$ odd with distinct eigenvalues and nullity one, the system \textup{(\ref{eq:1.1}) } is
completely integrable.
\end{corollary}
\begin{proof}
In this case we have $d=1$ and $n=2p+1$. All eigenvalues are distinct
with one of them being zero. The above proof of indepdence still 
holds, the only change being that the characteristic (and mininal)
polynomial of $N$ is of form $N w(N^2)$, where $w$ is 
a polynomiail of degree $(n-1)/2$.
\end{proof}

\section{Conclusion and Future work}
\label{Conclusion}

We have demonstrated integrability of the system \eqref{eq:1.1}
for appropriate $N$ by showing involution and independence of
a sufficient number of integrals.  
It is also of interest to analyze linearization on the Jacobi
variety of the curve 
\[
\operatorname{det}(zI- \lambda N - X) = 0
\]
using the theory discussed in \cite{AdvMVa2004} and 
\cite{Griffiths1985}, for example. We shall discuss these algebro-geometric aspects in a future paper. Independently \cite{LiTo2006} have shown the integrablity of the same system in precisely the two cases discussed in this paper employing different techniques; they use the loop group approach suggested by the Lax equation with parameter \eqref{eq:3.1} and give the solution in terms of factorization and the Riemann-Hilbert problem.

Another interesting variation of this system that we
shall consider in future work is the following.

\paragraph{A generalized system.}
The flow of  \eqref{eq:1.1} can be rendered more general by complexification.
Generalizing it to evolution in $\mathfrak{su}(n)$ yields an $n^2$-dimensional
flow of generalized rigid body type with two natural Hamiltonian
structures. Let $X_0\in\mathfrak{su}(n)$,
$N\in\operatorname{Sym}(n,\mathbb{R})$, and consider
\begin{equation}
  \label{eq:3.5}
   \dot{X} =[X^2,N]=[X,XN+NX], \qquad X(0)=X_0.
\end{equation}
Note that $X(t)$ evolves in $\mathfrak{su}(n)$ since one readily checks that
$[X,XN+NX] \in \mathfrak{su}(n) $.

Moreover, one can generalize this still further and take
$N\in\mathfrak{su}(n)$. 
We define
\begin{align*}
  H_1(X)&= \frac{1}{4} \operatorname{trace} X(XN+NX),\\
  H_2(X)&= \frac{1}{2} \operatorname{trace} X^2.
\end{align*}
Note that both Hamiltonians are real and that $H_2$ gives us our earlier
Hamiltonian in the case that $X$ is symmetric but that $H_1$ is 
zero in this case. 

\paragraph{Acknowledgments.} We thank G. Prasad for his observation regarding Lie
algebras. We also thank Percy Deift, Igor Dolgachev, Luc Haine,
 Rob Lazarsfeld,  
Alejandro Uribe, and Pol Vanhaecke for useful conversations that have helped us in the giving correct historical credits and whose suggestions improved the exposition.

\addtolength{\textheight}{-3cm}

\end{document}